# Development and cytotoxic response of two proliferative MDA-MB-231 and non-proliferative SUM1315 three-dimensional cell culture models of triple-negative basal-like breast cancer cell lines

**Dubois Clémence**[1,2], **Dufour Robin**[1,2], **Daumar Pierre**[1], **Aubel Corinne**[3], **Szczepaniak Claire**[4], **Blavignac Christelle**[4], **Mounetou Emmanuelle**[1,*], **Penault-Llorca Frédérique**[2,*] **and Mahchid Bamdad**[1,5]

[1]Université Clermont Auvergne, Institut Universitaire de Technologie, INSERM, U1240, Imagerie Moléculaire et Stratégies Théranostiques, F-63000 Clermont Ferrand, France

[2]Université Clermont Auvergne, Centre Jean Perrin, INSERM, U1240, Imagerie Moléculaire et Stratégies Théranostiques, F-63000 Clermont Ferrand, France

[3]Université Clermont Auvergne, Faculté de médecine, INSERM, U1240, Imagerie Moléculaire et Stratégies Théranostiques, F-63000 Clermont Ferrand, France

[4]Université Clermont Auvergne, Faculté de Médecine, Centre Imagerie Cellulaire Santé, F-63000 Clermont-Ferrand, France

[5]Current address: Institut Universitaire de Technologie de Clermont-Ferrand – Université Clermont Auvergne, Département Génie Biologique, Ensemble Universitaire des Cézeaux, CS 30086- 63172 AUBIERE CEDEX, France

[*]These authors have contributed equally to this work

*Correspondence to:* Mahchid Bamdad, *email:* mahchid.bamdad@uca.fr

*Keywords:* triple negative basal-like breast cancer cell lines, 3D cell culture, liquid overlay technique, cytotoxic/cytostatic drug activity

## ABSTRACT

Triple-Negative Basal-Like tumors, representing 15 to 20% of breast cancers, are very aggressive and with poor prognosis. Targeted therapies have been developed extensively in preclinical and clinical studies to open the way for new treatment strategies. The present study has focused on developing 3D cell cultures from SUM1315 and MDA-MB-231, two triple-negative basal-like (TNBL) breast cancer cell lines, using the liquid overlay technique. Extracellular matrix concentration, cell density, proliferation, cell viability, topology and ultrastructure parameters were determined. The results showed that for both cell lines, the best conditioning regimen for compact and homogeneous spheroid formation was to use 1000 cells per well and 2% Geltrex®. This conditioning regimen highlighted two 3D cell models: non-proliferative SUM1315 spheroids and proliferative MDA-MB-231 spheroids. In both cell lines, the comparison of 2D *vs* 3D cell culture viability in the presence of increasing concentrations of chemotherapeutic agents *i.e.* cisplatin, docetaxel and epirubicin, showed that spheroids were clearly less sensitive than monolayer cell cultures. Moreover, a proliferative or non-proliferative 3D cell line property would enable determination of cytotoxic and/or cytostatic drug activity. 3D cell culture could be an excellent tool in addition to the arsenal of techniques currently used in preclinical studies.

# INTRODUCTION

Triple-negative basal-like (TNBL) tumors comprise 70% of the basal-like tumor subtype and 15% of all breast cancers. They are characterized by the non-expression of estrogen receptors (ER) and progesterone receptors (PR), and the absence of HER2 over-expression or *ERBB-2* amplification [1–4]. TNBL tumors mainly affect young women and are frequently associated with hereditary predispositions (*BRCA1/2* germline mutations). This subtype has a very poor prognosis. TNBL tumours have a high proliferative capacity and may respond well to neoadjuvant chemotherapy or develop a resistance phenotype associated with metastases.

Conventional chemotherapy is based on different protocols such as FEC (5-fluorouracil, epirubicin, cyclophosphamide), FAC (5-fluorouracil, adriamycin, cyclophosphamide) or platinum salts (cisplatin), generally associated with side effects [5]. Otherwise, these tumors are not sensitive to classical breast targeted therapies since they do not express the relevant receptors (*i.e.* ER, PR and HER2). Thus, different groups have aimed to develop alternative targeted therapies. Targeting Epidermal Growth Factor Receptor 1 pathway with anti-EGFR Monoclonal antibodies (MoAb) or Tyrosine Kinase Inhibitors (TKI); or inhibiting the enzyme Poly-ADP-Ribose-Polymerase1 (reparation of single-strand breaks)(PARPi) have shown promising activities in this preclinical and clinical setting) [5–8].

Monolayer *in vitro* cell culture studies represent a gold standard high throughput screening for toxicity of chemotherapeutics. However, this type of culture does not reproduce the three-dimensional (3D) structural properties of tumors. In fact, these tumors are biochemically and structurally characterized by (i) the generation of hypoxic regions, (ii) intercellular interactions, (iii) nutrient and growth factor exchanges, and (iv) the production of extracellular matrix that is essential to tumor stability and regulation of cellular functions [9–11].

Therefore, over recent decades, 3D cell culture, mimicking the 3D organization of *in vivo*-like tumors, has regained interest. This innovative technique consists in forming aggregated and compact cell clusters called spheroids. In this conformation, cells gain different characteristics depending on their position [9]. Variable gradients develop for oxygen, nutrients, lactate accumulation and pH [9]. The easily accessible peripheral cells are metabolically active and able to proliferate. Conversely, cells in the centre have unfavourable conditions and ultimately become quiescent and form a necrotic core [12, 13]. Taken together these metabolic and biochemical heterogeneous features reproduce the inconstancy of tumors with respect to their response to treatment and predict more accurately the features of *in vivo* tumors [14]. Several 3D culture methods are available based on (i) the induction of mechanical forces *i.e.* centrifugation pellet culture, spinner flask culture and rotary cell culture systems, (ii) micromolding in hydrogels and (iii) gravity *i.e.* hanging drop culture and liquid overlay culture [13, 15, 16]. Amongst these methods, the liquid overlay technique uses cells capacity to aggregate by gravity on non-adherent surfaces and form spheroids within a short period of time.

The 3D cell culture model is being increasingly integrated into the drug development process as it appears to be a more accurate model when studying the impact of chemotherapeutics [17]. In fact, in terms of predicting potential drug efficacy, it is currently better than 2D cell culture [9].

This work focuses on the development of a 3D cell culture from two TNBL breast cancer cell lines, SUM1315 and MDA-MB-231, using the "liquid overlay" technique. For this, spheroid extracellular matrix concentration, cell density, metabolic activity, cell viability, proliferation, topology and ultrastructure parameters were firstly determined for both cell lines. Then, in order to better characterize 3D cell culture response, cell drugs sensibility of these spheroids was analysed in comparison to 2D cell culture.

# RESULTS

## 3D cell culture development from TNBL cell lines

3D cell culture development was carried out using two TNBL cell lines, SUM1315 and MDA-MB-231. For each cell line, four parameters were determined: (i) extracellular matrix protein concentration (ii) cell spheroid concentration, (iii) spheroid cell metabolic activity and (iv) spheroid cell viability/mortality.

## SUM1315 cell line

### Extracellular matrix protein concentration

For these experiments, Geltrex® was used as extracellular matrix proteins. SUM1315 cells were seeded at 5000 cells per well in "Ultra-Low-Attachment" microplates (Corning®) at "Day 0". Then, several conditioning regimens of Geltrex® were tested, *i.e.* just after cell seeding or 24h after cell seeding, and several Geltrex® concentrations (0.25 to 6%) were tested (Table 1, Figure 1B, 1C) and compared with a control cell culture without Geltrex® (Figure 1A).

The results showed that when Geltrex® was added just after cell seeding (at Day 0), loose cell aggregates were detected in the presence of 0.25 and 0.5% of Geltrex® concentrations (Table 1, Figure 1A). In the presence of 1 to 6% Geltrex®, only scattered spheroids were detected (Table 1, Figure 1B). When Geltrex® was added 24h after cell seeding (at Day 1), loose aggregates and scattered spheroids were observed in the presence of 0.25 to 0.5%

**Table 1: Extracellular matrix concentration determination for spheroid formation with the SUM1315 cell line**

| Percentage of Geltrex in the culture medium | Condition of Geltrex® adding | |
|---|---|---|
| | After cell seeding (Day 0) | 24h after cell seeding once aggregate formation (Day 1) |
| 0 (control) | Unique loose aggregate | Unique loose aggregate |
| 0.25 - 0.5 | Unique loose aggregate | Unique loose aggregate |
| 1 - 2 | Scattered spheroids | **Unique compact spheroid** |
| 3 - 4 - 5 - 6 | Scattered spheroids | Scattered spheroids |

Several Geltrex® concentrations from 0.25 to 6% were used. Two conditioning regimens of Geltrex® were tested: addition of Geltrex® (i) in the microplates just after cell seeding (corresponding to Day 0) and (ii) 24h after cell seeding and aggregate formation (corresponding to Day 1).

and 3 to 6% Geltrex® concentrations, respectively (Table 1, Figure 1A, 1B). In contrast, single compact spheroids were detected with 1 to 2% Geltrex® added 24h after cell seeding (Table 1, Figure 1C). Therefore 2% Geltrex® added at Day 1 was the conditioning regimen chosen for 3D culture development of the SUM1315 cell line.

**Cell spheroid concentration determination**

SUM1315 cell concentrations ranging from 50 to 10000 cells per well were cultured in the presence of 2% Geltrex® added after cell seeding at Day 1 (Figure 2A). Spheroid diameters were monitored every day from Day 0 (control) to Day 14 with light microscopy, using ToupView® software. Results showed that for 50, 2000 and 10000 cells per well, spheroid diameters were 299±54 µm, 1198±45 µm and 2149±50 µm at Day 0 and decreased to 89±16 µm, 320±21 µm and 570±21 µm at Day 1, respectively. SUM1315 spheroid diameters and shapes remained constant until Day 14 for concentrations of 200 to 2000 cells per well (313±14 µm for 2000 cells per well), and decreased for concentrations of 5000 and 10000 cells per well (401±31 µm for 10000 cells, $p<0.00001$) (Figure 2A, 2B).

**Metabolic activity study in 3D cell culture**

Metabolic activity was measured by the colorimetric resazurin test involving the reduction of resazurin into resorufin. For the SUM1315 cell line, this parameter was analysed at Day 1, Day 8 and Day 14 in spheroids of 50 to 10000 cells per well. Optic density (OD) of resorufin increased significantly for concentrations of 50, 200, 1000 and 2000 cells per well from Day 1 (with 0.006±0.002, 0.007±0.003, 0.051±0.008 and 0.120±0.044, respectively) to Day 8 (with 0.084±0.021 ($p<0.00001$), 0.076±0.022 ($p<0.001$), 0.179±0.015 ($p<0.00001$) and 0.233±0.043 ($p<0.01$, respectively)) (Figure 2B–2C). For cell concentrations of 5000 and 10000 per well, the OD of resorufin remained stable after 8 days of culture, with OD of 0.255±0.025 and 0.306±0.027 at Day 1 compared to 0.275±0.024 ($p=0.23$) and 0.324±0.028 ($p=0.73$) at Day 8, respectively.

After Day 8, the metabolic activity of SUM1315 spheroids dropped dramatically for all cell concentrations to Day 14 (OD of 0.047±0.010 for 5000 cells per well) (Figure 2–2C).

**Spheroid cell viability/mortality monitoring**

SUM1315 spheroid cell viability was analyzed during the experiment (14 days) by fluorescence microscopy using Live/Dead® kit (Molecular Probes). Cells forming spheroids remained viable (green markings) with all studied concentrations (50 to 10000 cells per well) for the 14 days of culture (Figure 2D), with no change in green fluorescent probe intensity between Day 8 and Day 14.

All these results showed that whatever the tested cell concentrations in the presence of 2% Geltrex® added after aggregate formation, compact and homogenous spheroids were formed. Moreover, spheroid diameters and viability remained stable for 14 days with all cell concentrations. Nevertheless, metabolic activity in spheroids decreased dramatically for all cell concentrations after 8 days of culture.

**MDA-MB-231 cell line**

**Extracellular matrix protein concentration**

For 3D cell culture development with the MDA-MB-231 cell line, the same Geltrex® concentrations and conditioning regimens were used as above. Single compact spheroids were detected with 2% Geltrex® added 24h after cell seeding (*data not shown*).

**Cell spheroid concentration determination**

For this cell line, as previously, spheroid diameters obtained with MDA-MB-231 cell concentrations from 50 to 10000 cells per well were measured every day from Day 0 to Day 14 with light microscopy using ToupView® Software (Figure 3A, 3B). At Day 0 for 50, 2000 and 10000 cells per well, diameters were 254±50 µm, 1203±100 µm and 2246±84 µm respectively. In contrast, spheroid diameters decreased significantly

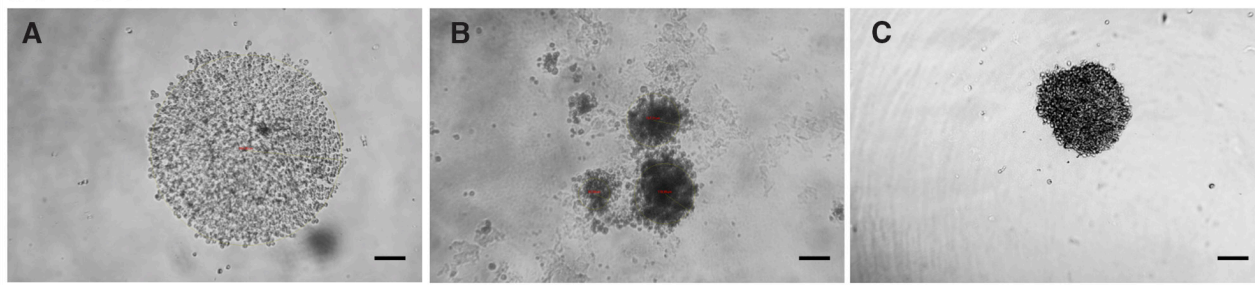

**Figure 1: Profile of spheroid formation with the SUM1315 cell line in light microscopy.** Aggregates and spheroids were observed by light microscopy at Day 1, using ToupView® software. Spheroid formation was considered when cells were compacted in a unique and opaque to light aggregate. Magnification = 100X, *scale bar = 200μm*. **(A)** SUM1315 loose cell aggregates, **(B)** scattered spheroids, **(C)** single compact spheroid.

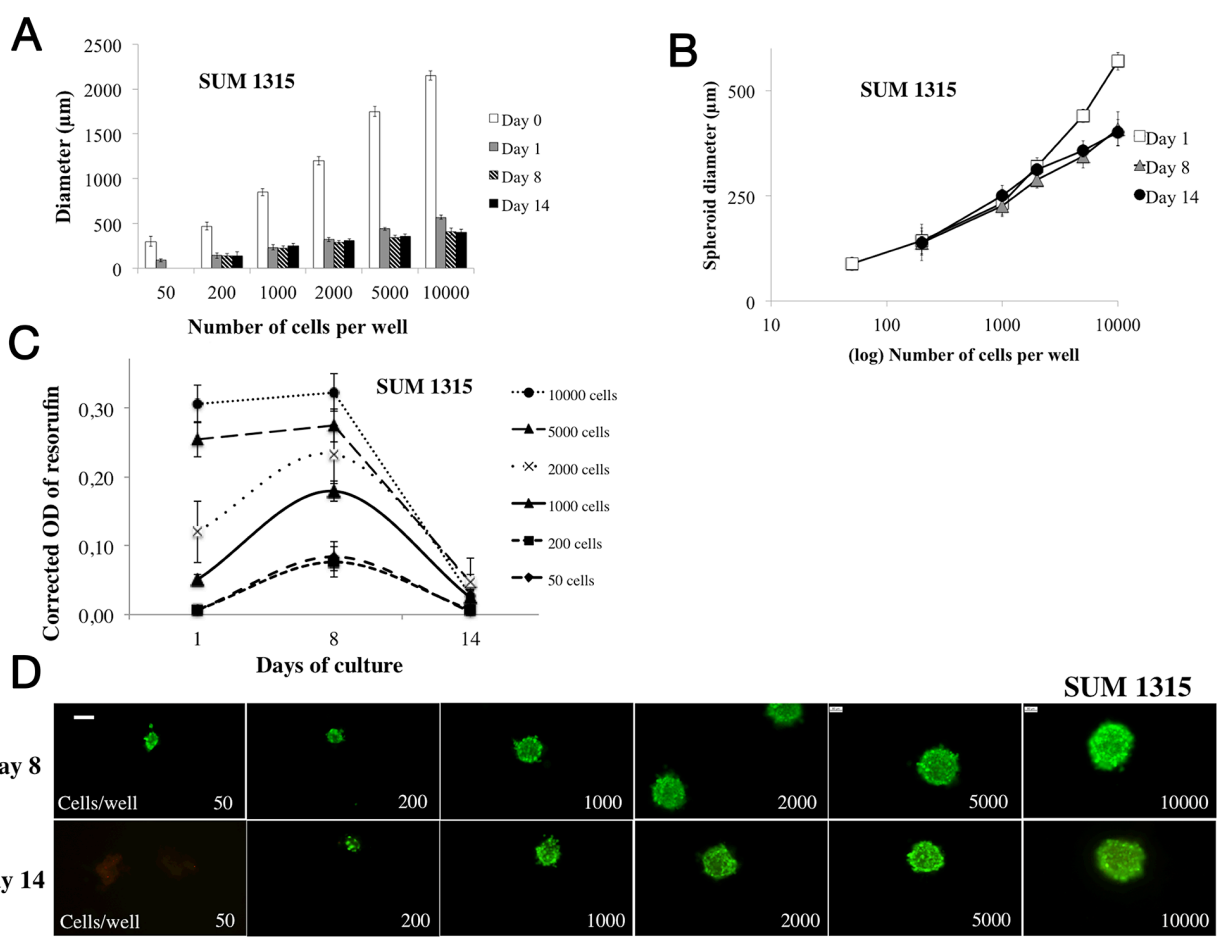

**Figure 2: Parameter determination for SUM1315 cell line spheroid formation.** For SUM1315 cell line spheroid formation, 50 to 10000 cells were seeded in 96-wells microplates ("ULA" for Ultra Low Attachment, Corning®) at Day 0 and 2% Geltrex® was added to the wells at Day 1. All experiments were carried out at Day 1, Day 8 and Day 14. **(A)** *Cell concentration determination for spheroid formation*: the diameter of each spheroid was measured (μm) by light microscopy with ToupView® software. Day 0 measurements correspond to aggregate diameters. **(B)** *Spheroid diameter evolution over time*: spheroid diameter (μm) was measured with ToupView software® for all tested cell concentrations (log10 scale). **(C)** *Cell metabolic activity in spheroids via resazurin test*: the *r*esazurin is reduced into resorufin by metabolically active cells. Corrected OD of resorufin (λ570- λ620 nm) was measured after 15h incubation with 60 μM resazurin in PBS. **(D)** *Cell viability/mortality in spheroids with LiveDead® imaging*: Green marking corresponds to calcein-AM penetration (viable cells). Red markings correspond to ethidium homodimer-1 cell penetration (dead cells). *Scale bar = 200 μm*. Graphs represent at least triplicate biological repeats and are displayed as mean ± SEM and *$p<0.05$, **$p<0.01$, ***$p<0.001$, ****$p<0.0001$ and *****$p<0.00001$.

after adding Geltrex®. At Day 1, diameters were 134±23 µm (p<0.00001), 612±59 µm (p<0.00001) and 826±90 µm (p<0.00001) for 50, 2000 and 10000 cells per well, respectively. Furthermore, the monitoring of MDA-MB-231 spheroids showed a significant increase in diameters at Day 8 for concentrations of 50 to 2000 cells per well with 317±65 µm (p<0.00001) and 843±69 µm (p<0.00001), respectively. Moreover, for concentrations of 5000 and 10000 cells per well, no diameter increase was detected under these experimental conditions (814±67 µm for 10000 cells per well at Day 8, p=0.68 and 848±76 µm at Day 14, p=0.27) (Figure 3A, 3B). After 14 days, spheroid diameters continued increasing significantly for cell concentrations of 50 to 1000 cells per well (608±148 µm for 50 cells per well, p<0.00001).

**Metabolic activity study in 3D cell culture**

Metabolic activity assessment was carried out for 14 days, in MDA-MB-231 spheroids seeded with 50 to 10000 cells per well (Figure 3C). The OD of resorufin significantly increased over time with all cell concentrations after 8 days of culture. At Day 1, OD was 0.010±0.003, 0.062±0.005 and 0.165±0.023 and increased to 0.054±0.006 (p<0.00001), 0.115±0.010 (p<0.00001) and 0.244±0.035 (p<0.001) at Day 8 for 200, 1000 and 10000 cells per well respectively.

The metabolic activity of spheroids continued increasing for spheroids of 50 to 200 cells per well (0.167±0.007 for 200 cells, p<0.00001), remained stable for spheroids of 1000 cells per well (0.128±0.014, p=0.11), and decreased significantly for spheroids of 2000, 5000 and 10000 cells per well (0.074±0.037 for 10000 cells, p<0.00001 compared to Day 8).

**Spheroid cell viability/mortality monitoring**

MDA-MB-231 spheroid cell viability was monitored during the experiment (14 days) by fluorescence microscopy using Live/Dead® kit (Molecular Probes) (Figure 3D). Cells forming spheroids remained viable (green markings) with all studied concentrations (50 to 10000 cells per well) for the 8 days of culture. In contrast, at day 14, green markings were less intense (in same conditions of exposure and gamma) with spheroids formed with 2000 to 5000 cells per well, and even absent with 10000 cells per well.

Overall results obtained with the MDA-MB-231 cell line also showed that whatever the tested cell concentrations in the presence of 2% Geltrex® added after aggregate formation, compact and homogenous spheroids were formed. Spheroid diameters increased clearly for concentrations of 50 to 2000 cells per well for 14 days, suggesting a proliferative property of the MDA-MB-231 cell line under these experimental conditions. Moreover, metabolic activity in spheroids increased clearly during 8 days of culture presenting stable viability for all cell concentrations. Nevertheless, for cell concentrations superior to 2000 per well, metabolic activity and viability (Live/Dead® kit) dropped after 14 days.

For further experiments, with both SUM1315 and MDA-MB-231 spheroid cell cultures, the conditioning regimen of 1000 cells per well and 2% Geltrex® were chosen.

**Topological and ultrastructural 3D cell culture characterization**

**3D cell culture architecture of SUM1315 and MDA-MB-231 spheroids with Scanning Electron Microscopy**

Scanning Electron Microscopy (SEM) was used to assess the 3D structure of cells forming the spheroids. Both SUM1315 (Figure 4A, 4B, 4C) and MDA-MB-231 (Figure 4D, 4E, 4F) compact and homogenous spheroids exhibited rounded shapes of about 300 µm with granular surfaces covered with stacked cells. These cells in clusters were 10 µm in diameter and most of them presented rounded morphology. They are all closely juxtaposed with extracellular matrix (ECM) filling the empty spaces.

**Ultrastructure of SUM1315 and MDA-MB-231 spheroids with Transmission Electron Microscopy**

The organization and ultrastructure of cells forming the spheroids were analyzed using Transmission Electron Microscopy (TEM) of SUM1315 (Figure 5A, 5B, 5C, 5D) and MDA-MB-231 (Figure 5E, 5F, 5G, 5H) spheroids. SUM1315 (Figure 5A) and MDA-MB-231 spheroids (Figure 5E) displayed adjoined cells with intact plasma and nuclear membranes. Cells established contact by two types of cell junctions: tight junctions (Figure 5C, 5G) and anchoring junctions (*zonula adherens*, Figure 5D) as well as desmosomes (Figure 5B, 5H). Ultrastructural analysis at higher magnifications (M=15000 – 20000 X) revealed the presence of conventional organelles such as mitochondria, rough endoplasmic reticulum and lysosomes (Figure 5B, 5F).

**2D and 3D TNBL cell culture metabolic activity comparison**

The comparison of metabolic activity between 2D and 3D cell cultures (as described above) was analyzed with the resazurin test after 5 days of culture of SUM1315 (Figure 6A) and MDA-MB-231 (Figure 6B) cell lines seeded at 1000 cells/well. SUM1315 cell metabolic activity in 2D cell culture was 0.098±0.005 AU and significantly higher than 3D cell culture with 0.053±0.008 AU (p<0.00001) (Figure 6A). Similarly, for the MDA-MB-231 cell line, it was 0.179±0.025 AU in 2D and significantly higher than 3D cell culture with 0.077±0.013 AU (p<0.00001) (Figure 6B). These results showed that metabolic activity of cells cultured in 3D was lower than cells cultured in monolayer under these experimental conditions.

**Analysis of 2D and 3D TNBL cell culture sensitivity to drugs**

Sensitivity of SUM1315 and MDA-MB-231 cell lines in 2D and 3D cell cultures (as described above) was studied in the presence of three conventional chemotherapeutic drugs *i.e.* cisplatin, docetaxel and epirubicin. A quantitative resazurin test was used after 5 days of treatment to determine cell viability (Figures 7, 8, 9). The percentage of cell viability was calculated by the ratio of the quantity of resorufin formed by treated cells in comparison with untreated cells. For 3D cell culture, spheroid diameter measurements and qualitative viability/mortality analysis using the Live/Dead® kit were also performed at Day 5. In parallel, controls and solvent controls with DMSO 0.1% were also performed under the same culture conditions.

**Sensitivity to cisplatin**

For the SUM1315 cell line, analysis of the ratio of the quantity of resorufin formed in DMSO 0.1% treated cells in comparison with untreated control cells showed no impact of DMSO on cell viability, with 99±11% for 2D cell culture (p=0.69) and 101±24% for 3D cell culture (p=0.61). With cisplatin, the analysis of cell viability in 3D and 2D cell cultures at day 5 was of 80±14% *vs* 101±8% with 0.01 μM (p<0.00001), 72±9% *vs* 104±5% with 0.1 μM (p<0.00001), 74±9% *vs* 103±4% with

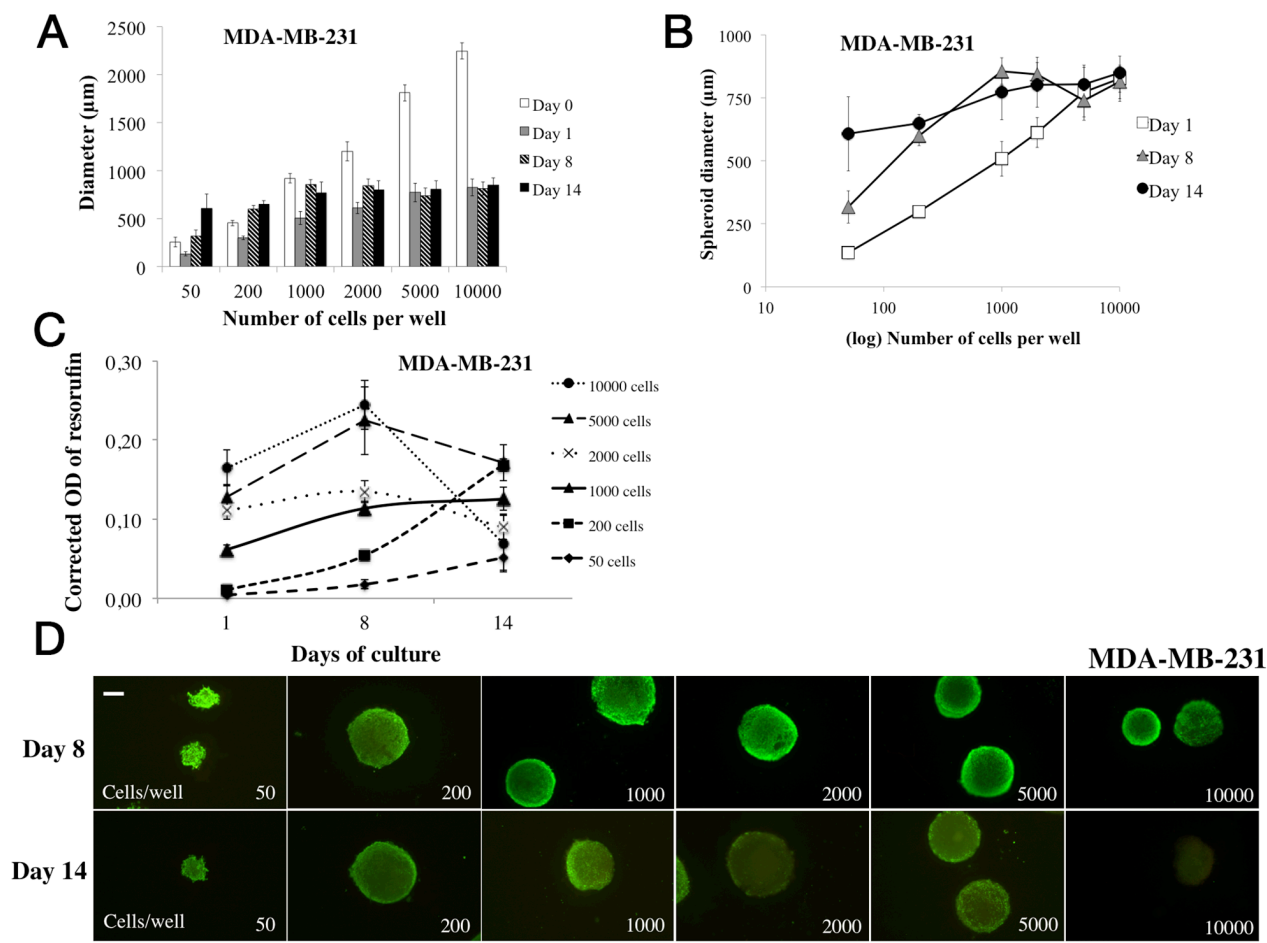

**Figure 3: Parameter determination for MDA-MB-231 cell line spheroid formation.** For MDA-MB-231 cell line spheroid formation, 50 to 10000 cells were seeded in 96-wells microplates ("ULA" for Ultra Low Attachment, Corning®) at Day 0 and 2% Geltrex® was added to the wells at Day 1. All experiments were carried out at Day 1, Day 8 and Day 14. **(A)** *Cell concentration determination for spheroid formation*: the diameter of each spheroid was measured by light microscopy with ToupView® software. Day 0 measures correspond to aggregate diameters. **(B)** *Spheroid diameter evolution over time*: spheroid diameter (μm) was measured with ToupView software® for all tested cell concentrations (log10 scale). **(C)** *Cell metabolic activity in spheroids via resazurin test*: the *r*esazurin is reduced into resorufin by metabolically active cells. Corrected OD of resorufin (λ570- λ620 nm) was measured after 15h incubation with 60 μM resazurin in PBS. **(D)** *Cell viability/mortality in spheroids with Live/Dead® imaging*: Green marking corresponds to calcein-AM penetration (viable cells). Red markings correspond to ethidium homodimer-1 cell penetration (dead cells). *Scale bar = 200 μm*. Graphs represent at least triplicate biological repeats and are displayed as mean ± SEM and *p<0.05, **p<0.01, ***p<0.001, ****p<0.0001 and *****p<0.00001.

1 μM (p<0.00001), and 66±4% *vs* 54±6% with 10 μM (p<0.00001), respectively (Figure 7Aa). These results showed a slight decrease in 3D SUM1315 cell viability whatever the cisplatin concentrations. In contrast, viability of 2D cell cultures decreased significantly only with a high cisplatin concentration (10 μM).

SUM1315 spheroid diameters were then measured with ToupView® software. The results showed similar diameters between 0.01 and 0.1 μM cisplatin treated spheroids (264±10 μm, p=0.14 and 264±12 μm, p=0.17, respectively) compared to controls (267±11 μm) after 5 days of treatment (Figure 7Ab). Nevertheless, a decrease in spheroid diameters was detected after treatment with 1 and 10 μM cisplatin (223±11 μm, p<0.00001 and 204±12 μm, p<0.00001, respectively). Live/Dead® kit analysis of viability and mortality of 3D cell cultures showed similar cell viability of spheroids treated with 0.01 to 10 μM cisplatin to control spheroids (Figure 7Ac).

For the MDA-MB-231 cell line, analysis of the ratio of the quantity of resorufin formed in DMSO 0.1% treated

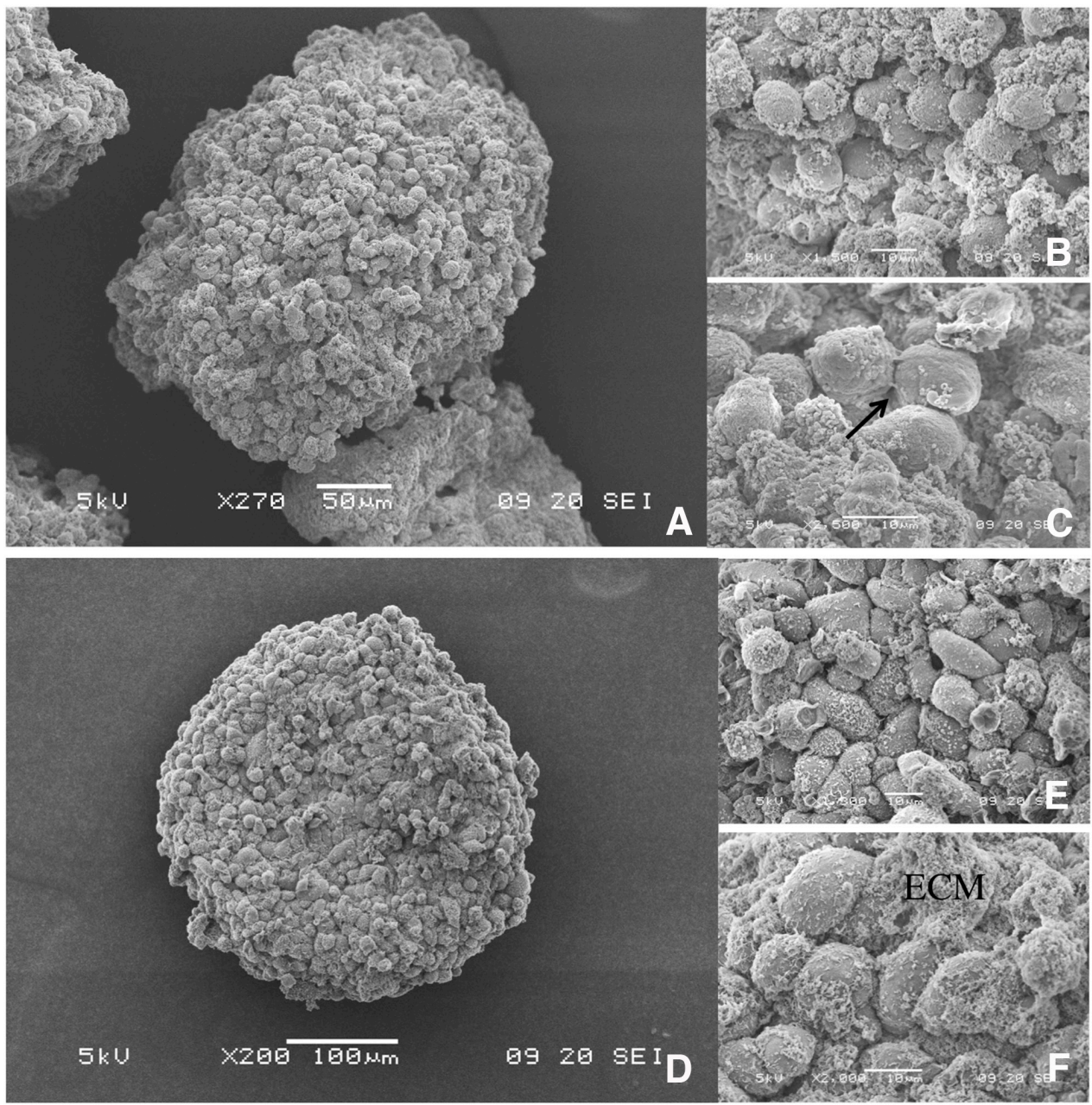

**Figure 4: 3D topology of SUM1315 and MDA-MB-231 spheroids by Scanning Electron Microscopy (SEM).** Cells were seeded at 1000 cells per well with 2% Geltrex®, added at Day 1. SEM images were taken at Day 5 with Jeol 6060 LV scanning electron microscope. Scale bars are shown on all images. *M=magnification* SUM1315 spheroids: **(A)** *M= 270X,* **(B)** *M=1500X,* **(C)** *M=2500X* MDA-MB-231 spheroids at **(D)** M=200X, **(E)** M=1300X, **(F)** M=2000X. ECM= extracellular matrix, Black arrow= cell junction

www.impactjournals.com/oncotarget                    7                    Oncotarget

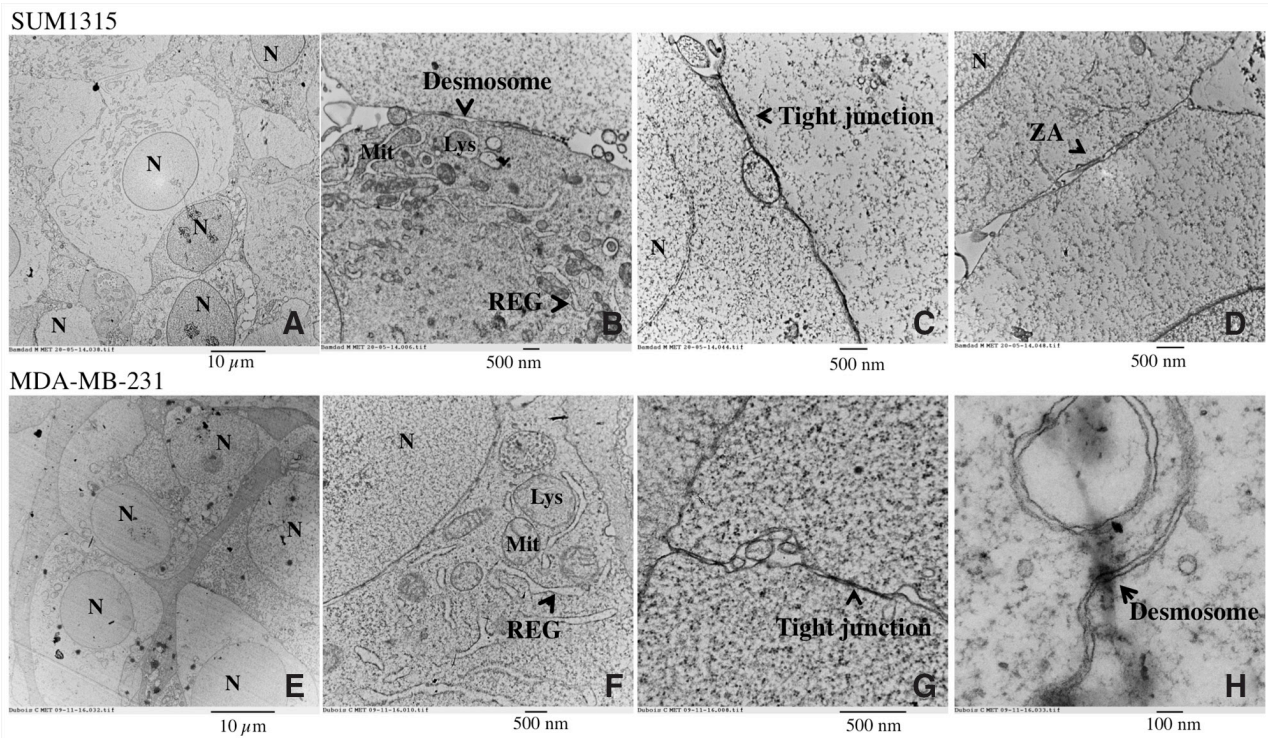

**Figure 5: Ultrastructure of SUM1315 and MDA-MB-231 spheroids by Transmission Electron Microscopy (TEM).** Cells were seeded at 1000 cells per well with 2% Geltrex®, added at Day 1. Images were taken after 5 days of culture with transmission electron microscope Hitachi H-7650. Scale bars are shown on all images. *M=magnification* SUM1315 cells ultrastructure in 3D cell culture conditions: **(A)** *M=3000X,* **(B)** *M=15000X,* **(C)** *M=25000X,* **(D)** *M=25000X.* MDA-MB-231 cells ultrastructure in 3D cell culture conditions: **(E)** *M=3000X,* **(F)** *M=20000X,* **(G)** *M=60000X,* **(H)** *M=120000X.* N=nucleus, Mit=mitochondria, Lys=lysosome, REG= Rough Endoplasmic Reticulum, ZA=zonula adherens. Scale bars are shown on all images.

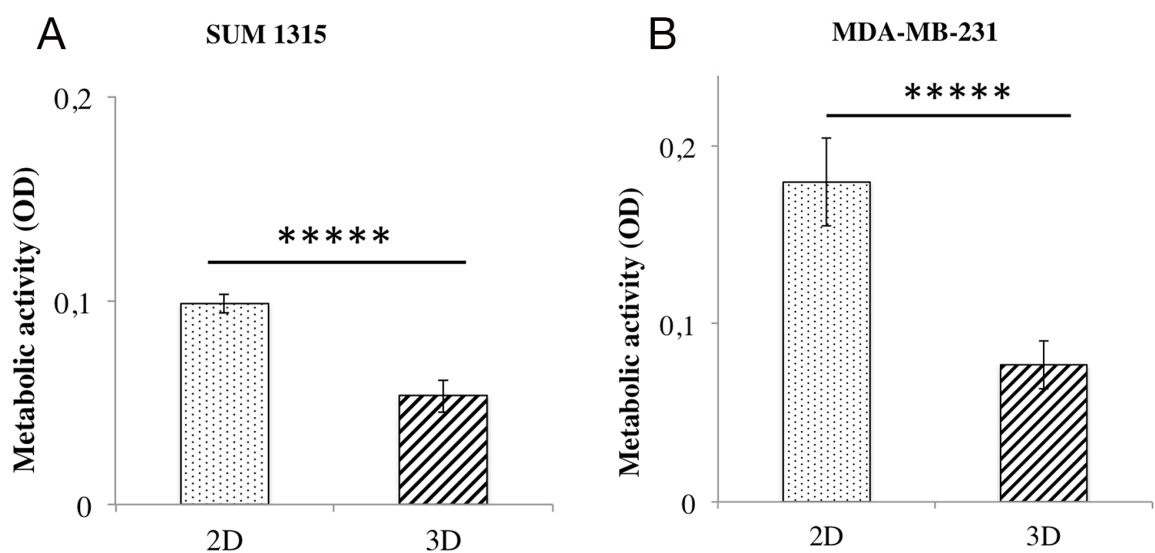

**Figure 6: Metabolic activity comparison of 2D *vs* 3D cell cultures with resazurin test.** The resazurin is reduced into resorufin by metabolically active cells. The amount of resorufin formed thus reflects the metabolic activity of the cells. Corrected OD of resorufin ($\lambda$570-$\lambda$620 nm) was measured after 6h of incubation with resazurin in PBS (60 μM) in the presence of 1000 cells per well at seeding for both cell lines and both culture conditions. Measures were made for both cell lines **(A)** SUM1315 and **(B)** MDA-MB-231 and all conditioning regimens after one day of culture. Graphs represent at least triplicate biological repeats and are displayed as mean ± SEM, ***** = $p<0.00001$.

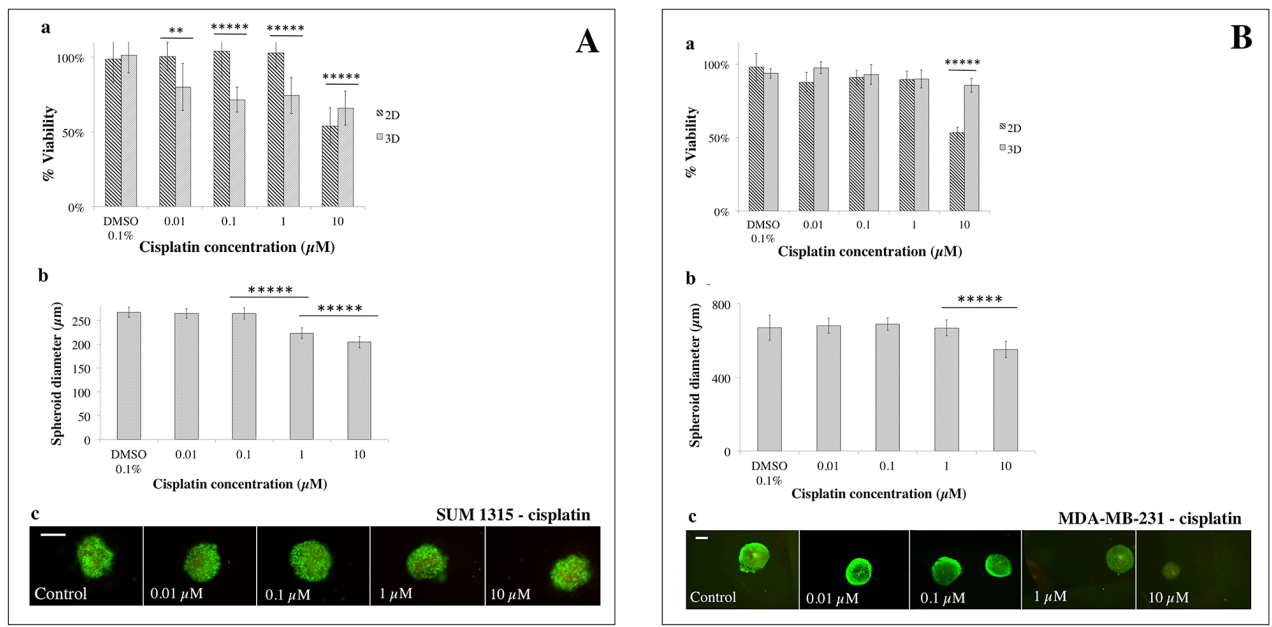

**Figure 7: 2D *vs* 3D cell culture sensitivity to cisplatin. (A)** SUM1315 and **(B)** MDA-MB-231 cell line sensitivity to cisplatin (0.01, 0.1, 1, 10 μM) was assessed in 2D and 3D cell culture conditions after 5 days of treatment. **(a)** *Cell viability in 2D and 3D culture conditions with the resazurin test:* viability was calculated by the ratio of OD of resorufin formed in cisplatin-treated cells to 0.1% control DMSO cells. **(b)** *Spheroid diameter measurement:* this parameter was measured with ToupView® software (μm). **(c)** *Live/Dead® spheroid imaging*: Green marking corresponds to calcein-AM penetration (viable cells). Red markings correspond to ethidium homodimer-1 cell penetration (dead cells). *Scale bar =200 μm.* Results are displayed as mean ± SEM where $*p<0.05$, $**p<0.01$, $***p<0.001$, $****p<0.0001$ and $*****p<0.00001$.

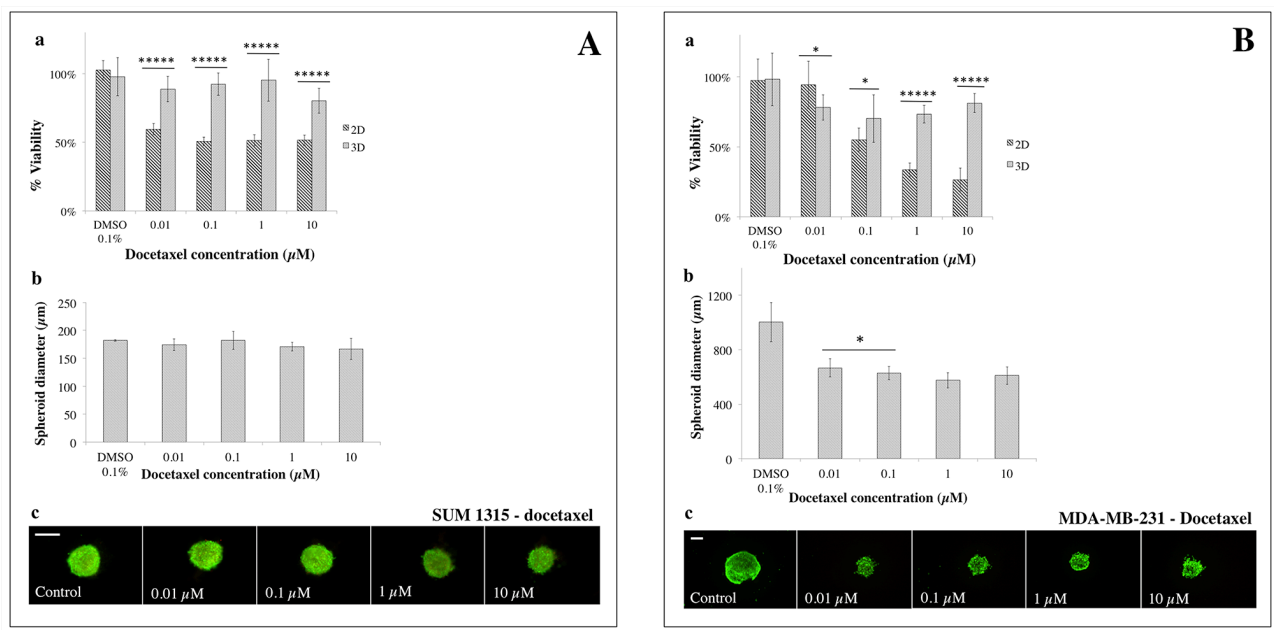

**Figure 8: 2D *vs* 3D cell culture sensitivity to docetaxel. (A)** SUM1315 and **(B)** MDA-MB-231 cell lines sensibility to docetaxel (0.01, 0.1, 1, 10 μM) was assessed in 2D and 3D cell culture conditions after 5 days of treatment. **(a)** *Cell viability in 2D and 3D culture conditions with the resazurin test:* viability was calculated by the ratio of OD of resorufin on docetaxel-treated cells to 0.1% control DMSO cells. **(b)** *Spheroid diameter measurement:* this parameter was measured with ToupView® software (μm). **(c)** *Live/Dead® spheroid imaging*: Green marking corresponds to calcein-AM penetration (viable cells). Red markings correspond to ethidium homodimer-1 cell penetration (dead cells). *Scale bar =200 μm.* Results are displayed as mean ± SEM where $*p<0.05$, $**p<0.01$, $***p<0.001$, $****p<0.0001$ and $*****p<0.00001$.

cells in comparison with untreated control cells showed no impact of DMSO at this concentration, with 98±5% for 2D (p=0.21) and 94±7% for 3D cell culture (p=0.16). With cisplatin, the analysis of 3D and 2D cell viability after treatment showed 97±8% *vs* 88±5% viability after 0.01 μM cisplatin (p<0.01), 93±7% *vs* 91±15% for 0.1 μM (p=0.68), 90±5% *vs* 89±16% for 1 μM (p=0.94) and 86±4% *vs* 53±4% for 10 μM (p<0.00001), respectively (Figure 7Ba).

These results showed slight cisplatin toxicity with 10 μM on MDA-MB-231 3D cell culture compared to 2D. Moreover, $IC_{50}$ could not be determined for either cell culture.

MDA-MB-231 spheroid diameters analysis at Day 5, in the presence of different concentrations of cisplatin showed similar diameters to controls after treatment with 0.01 to 1 μM cisplatin. Nevertheless, a significant decrease in spheroid size with 10 μM (551±33 μm) compared to controls (680±39 μm, p<0.00001) (Figure 7Bb) was detected. In parallel, similar green markings were observed in spheroids treated with 0.01 to 1 μM cisplatin compared to controls, and lower viability was detected for spheroids treated with 10 μM (Figure 7Bc).

**Sensitivity to docetaxel**

For the SUM1315 cell line, the analysis of the ratio of the quantity of resorufin formed in DMSO 0.1% treated cells in comparison with untreated control cells showed no toxicity of DMSO at this concentration, with 102±% for 2D (p=0.27) and 98±14% for 3D cell cultures (p=0.73).

SUM1315 cell line sensitivity to docetaxel was assessed under the same culture conditions (Figure 8A). 3D and 2D cell culture viability was 89±9.0% *vs* 58±4% with 0.01 μM docetaxel (p<0.00001), 92±8% *vs* 49±3.0% for 0.1 μM (p<0.00001), 95±15% *vs* 50±4% for 1 μM (p<0.00001) and 80±9.0% *vs* 50±4% for 10 μM (p<0.00001), respectively (Figure 8Aa). These results showed that sensitivity of the SUM1315 2D cell line to docetaxel remained relatively stable during the experiment in the presence of increasing concentrations of the drug, with the $IC_{50}$ value never reached. The same results were observed for the SUM1315 3D cell line. Nevertheless, as with cisplatin, SUM1315 3D cell culture was less sensitive to docetaxel than the 2D cell culture (p<0.00001 for all tested concentrations).

SUM1315 spheroid diameters remained stable in the presence of increasing concentrations of docetaxel (175±11 μm, 182±16 μm, 171±8 μm, 167±19 μm for respectively 0.01, 0.1, 1, 10 μM docetaxel, compared to controls 182±1 μm (Figure 8Ab). Live/Dead® kit analysis of viability and mortality of 3D cell cultures showed similar cell viability of treated spheroids to controls for all docetaxel concentrations (Figure 8Ac). These results correlated with the high viability rates observed with the resazurin test for 3D cell culture.

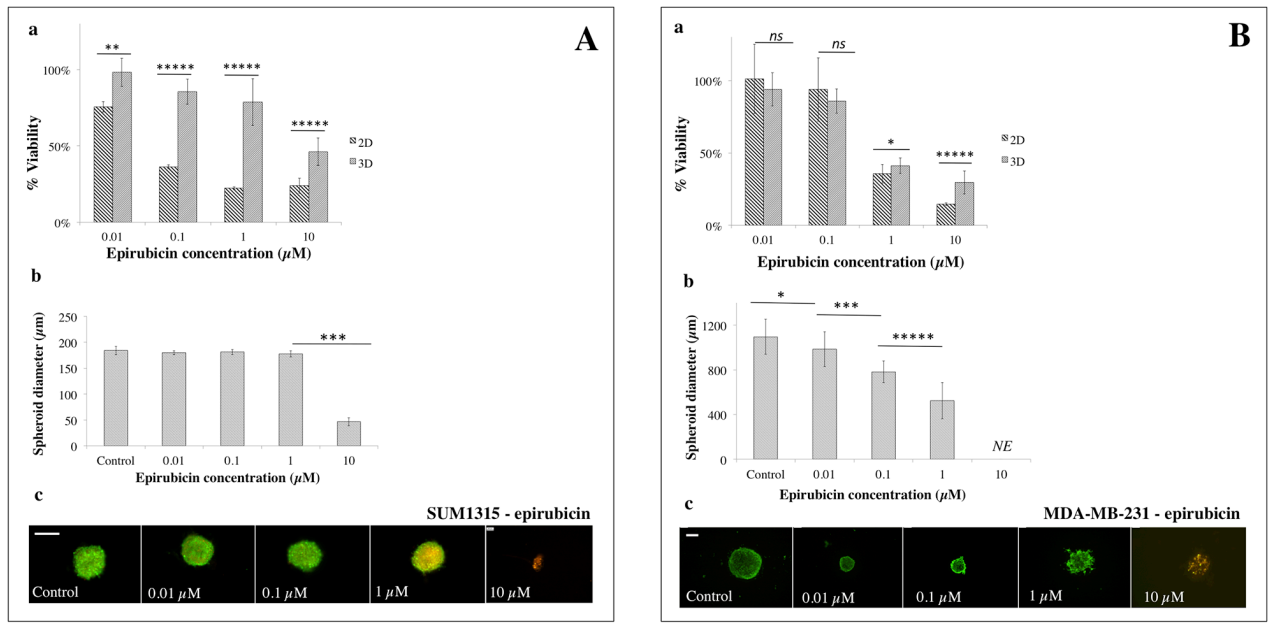

**Figure 9: 2D *vs* 3D cell culture sensitivity to epirubicin.** **(A)** SUM1315 and **(B)** MDA-MB-231 cell line sensitivity to epirubicin (0.01, 0.1, 1, 10 μM) was assessed in 2D and 3D cell culture conditions after 5 days of treatment. **(a)** *Cell viability in 2D and 3D culture conditions with the resazurin test:* viability was calculated by the ratio of OD of resorufin in epirubicin-treated cells to 0.1% control DMSO cells. **(b)** *Spheroid diameter measurement:* this parameter was measured with ToupView® software (μm). **(c)** *Live/Dead® spheroid imaging*: Green marking corresponds to calcein-AM penetration (viable cells). Red markings correspond to ethidium homodimer-1 cell penetration (dead cells). *Scale bar =200 μm.* Results are displayed as mean ± SEM where *p<0.05, **p<0.01, ***p<0.001, ****p<0.0001 and *****p<0.00001.

For the MDA-MB-231 cell line, the analysis of the ratio of the quantity of resorufin formed in DMSO 0.1% treated cells in comparison with untreated control cells showed no toxicity of DMSO at this concentration, with 97±15% for 2D (p=0.65) and 98±19% for 3D cell culture (p=0.80).

With docetaxel, 3D *vs* 2D cell culture viability was 78±9% *vs* 94±17% for 0.01 μM (p<0.05), 70±17% *vs* 55±8% for 0.1 μM (p<0.05), 73±6% *vs* 33±5% for 1 μM (p<0.00001), and 81±7% *vs* 27±8% for 10 μM (p<0.00001), respectively (Figure 8Ba). MDA-MB-231 3D cell culture was again less sensitive than 2D cell culture with $IC_{50}$ values of 0.2±0.004 μM for 2D and superior to 10 μM for 3D cell culture, and a ratio 3D/2D > 58 fold.

After docetaxel treatment, MDA-MB-231 spheroid diameters decreased significantly to a plateau with all tested concentrations (666±66 μm with 0.01 μM and 609±64 μm with 10 μM) in comparison with controls (1002±144 μm, p<0.00001)(Figure 8Bb). Live/Dead® kit analysis showed that MDA-MB-231 spheroid viability remained stable in the presence of increasing concentrations of docetaxel (Figure 8Bc).

**Sensitivity to epirubicin**

SUM1315 3D and 2D cell culture viability after epirubicin treatment was 98±12% *vs* 76±3% for 0.01 μM (p<0.001), 85±15% *vs* 36±1% for 0.1 μM (p<0.00001), 46±4% *vs* 22±1% for 1 μM (p<0.00001) and 46±4% *vs* 24±5% for 10 μM (p<0.00001), respectively (Figure 9Aa). $IC_{50}$ values of 2D and 3D cell culture were 0.1±0.004 μM against 7.6±0.5 μM, respectively. These results showed that 3D cell culture was 140 fold less sensitive to epirubicin than 2D monolayer cell culture.

SUM1315 spheroid diameters remained stable in the presence of 0.01, 0.1 or 1 μM epirubicin (180±4 μm, 181±5 μm and 178±6 μm, respectively) compared to controls (184±8 μm) (Figure 9Ab). However, spheroid diameters decreased significantly (47±8 μm, p<0.001) with 10 μM of the drug. Live/Dead® kit analysis confirmed that SUM1315 spheroids treated with 0.01 and 0.1 μM showed the same viability as controls. In contrast, spheroids treated with 1 μM epirubicin presented a necrotic core (yellow markings), and were dead with 10 μM (red markings) (Figure 9Ac).

For the MDA-MB-231 cell line, 3D and 2D cell culture viability was 94±11% *vs* 101±24% for 0.01 μM epirubicin (p=0.29), 86±8% *vs* 94±22% for 0.1 μM (p=0.19), 41±5% *vs* 36±7% for 1 μM (p<0.05) and 30±8% *vs* 15±1% for 10 μM (p<0.00001), respectively (Figure 9Ba). $IC_{50}$ values were 0.6±0.003 μM for 2D and 0.6±0.1 μM for 3D cell culture. These results showed that MDA-MB-231 3D cell culture was as sensitive to epirubicin as 2D cell culture.

MDA-MB-231 spheroid diameters decreased significantly in a dose-dependent manner, in the presence of increasing concentrations of epirubicin: 986±155 μm (p<0.05), 782±97 μm (p<0.00001) and 525±163 μm (p<0.00001) for 0.01, 0.1 and 1 μM epirubicin, respectively, against 1096±157 μm for untreated spheroids (Figure 9Bb). With 10 μM epirubicin, spheroids were destroyed. Live/Dead kit analysis of viability and mortality showed that for epirubicin concentrations of 0.01 to 1 μM, spheroid viability remained stable (Figure 9Bc). In contrast, remaining spheroids treated with 10 μM exhibited a predominance of diffuse red marked cells (Figure 9Bc).

## DISCUSSION

To optimize *in vitro* assays for the evaluation of drug sensitivity in TNBL breast cancer cell lines, we have conducted different studies to develop optimal 3D cell cultures from SUM1315 and MDA-MB-231 cell lines, with the "liquid overlay" technique ("Ultra-Low-Attachment" microplates, Corning®). Indeed, this technique presents several benefits, such as a single-per-well spheroid formation, generated naturally by gravity. Futhermore, this method has been validated for a large number of cancer cell lines [18, 19]. The works allowed for determination of optimal culture conditions, metabolic activity, morphology and ultrastructure of cells within the spheroids. After model developing, the impact of several anticancer agents was studied on both 3D cell cultures in comparison to 2D cell cultures sensitivity.

For the development of 3D compact cell mass cultures with SUM1315 and MDA-MB-231 cell lines, the need for an extracellular matrix was first investigated. For this, Geltrex™ LDEV-Free (Reduced Growth Factor Basement Membrane Matrix, GIBCO®) was used as a specific mixture matrix as it contains mainly laminin, collagen IV, entactin and heparin sulfate proteoglycan. With both cell lines, the 2% Geltrex® concentration resulted in the formation of a single compact spheroid per well, with a 3.6±0.3 fold cell compaction for SUM1315 cell line and a 2.1±0.4 fold compaction for the MDA-MB-231 cell line, whatever cell density. Indeed, some cell lines require specific components mimicking the extracellular matrix (ECM) such as laminin or collagen IV, in their culture medium. These components are responsible for the cell/extracellular matrix connections, and are able to generate an environment leading to the formation of intercellular connections [10, 15, 20]. Moreover, the Triple-Negative Basal-Like SUM1315 and MDA-MB-231 cell lines express low protein intercellular junctions such as E-cadherin, which limit the ability of cells to organize themselves spontaneously into compact spheroids [19–22]. Similar Geltrex® concentrations and conditioning regimens were required to form spheroids with MDA-MB-231 in the literature [11, 15].

Cell density for compact and homogenous spheroid formation was then assessed with both cell lines under the same experimental conditions as described above. Overall results showed different spheroid diameter evolutions according to TNBL cell line, under the same experimental conditions. Indeed, SUM1315 spheroids did not evolve over time in comparison to MDA-MB-231 spheroids which had expanded. This could be explained by a proliferative capacity of the MDA-MB-231 cell line under these experimental conditions in comparison to the SUM1315 cell line that may exhibit a "stem-cell like" phenotype with slowed down proliferation rate [23]. In 3D culture models, cells forming spheroids may or may not proliferate according to cell type and the extracellular matrix environment [24, 25]. Nevertheless, in both TNBL 3D cell cultures, spheroid diameters did not exceed 1000 μm, which could be explained by the fact that the spheroids had reached their growth plateau [19].

The metabolic activity of 3D-non-proliferative SUM1315 and 3D-proliferative MDA-MB-231 spheroids was then measured at day 1, day 8 and day 14 using the colorimetric resazurin test. This test is safe for cells, easily penetrates spheroids and has already been validated for a large number of 3D cell culture models [26, 27]. Overall results showed that SUM1315 spheroids exhibited metabolic activity only until 8 days of culture, whatever the cell density, whereas MDA-MB-231 spheroids exhibited metabolic activity until 14 days of culture with low cell concentrations. The decrease in metabolic activity and/or cell viability with LiveDead® Kit, detected in the presence of high cell concentrations and/or long-term incubation conditions, may be caused by the formation of necrotic or quiescent areas in spheroids, already described in other 3D models [19].

Optimal 3D cell culture conditions with both non-proliferative SUM1315 and proliferative MDA-MB-231 cell lines required adding 2% Geltrex® after aggregate formation in "ULA microplates" (Corning®) in the presence of 1000 cells per well. This conditioning regimen of 3D culture was used to analyze the topology of spheroids with both cell lines by first Scanning Electron Microscopy (SEM). SEM observations showed the homogenous three-dimensional topology of these masses with tightly packed cells at the surface, covered with extracellular matrix. Then, Transmission Electron Microscopy (TEM) observations of "in-depth" spheroid profiles demonstrated that both cell lines resulted in the formation of compact and organized aggregates. Indeed, these spheroids exhibited round shapes with juxtaposed cells presenting intact membranes, intracellular organelles and intercellular junctions [27].

For 3D cell culture development, it was necessary to compare spheroids metabolic activity to 2D monolayer cultures. For both cell lines under the same experimental conditions, metabolic activity of 2D cell culture was significantly higher than that of 3D cell culture, probably due to cellular heterogeneity in spheroids also described in literature [14, 27]. Moreover, the comparison of metabolic activity between both cell lines in 2D or 3D cell cultures showed systematic lower activity of SUM1315 than MDA-MB-231 cell line, probably correlated to their respective non-proliferative/proliferative profiles.

All these results showed the development of two TNBL 3D cell culture models using the "liquid overlay" technique that allowed the formation of single-per-well spheroids with quick and easy handling, presenting homogenous shapes with high reproducibility. These two 3D non-proliferative and proliferative models were then used to study anticancer agents effectiveness in comparison to 2D cell culture.

For these experiments, three drugs used in treatment protocols for BLTN breast cancers, alone or in combination, such as cisplatin, docetaxel and epirubicin were chosen [28]. Quantitative viability tests, spheroid diameter and Live/Dead® viability tests were performed. Overall results showed that with both non-proliferative SUM1315 and proliferative MDA-MB-231 TNBL cell lines, sensitivity profiles of 3D and 2D cell cultures seemed to be cell line- and drug-dependent. Indeed, cisplatin (platinum salt) is a cell cycle-dependent alkylating agent which creates inter and intrastrand DNA links, inducing serious DNA double-strand damage after the replication fork. Hence, this class of drug is promoted for BLTN tumors in combination with other chemotherapy agents [28]. Similarly, docetaxel is an antimitotic agent from the taxane family which stops cell mitosis division in proliferating tumoral cells. It has shown to be of interest for TNBL tumor treatment when used in combination with anti-angiogenic agents [28]. In our experimental conditions, cisplatin and docetaxel effectiveness analysis showed cytostatic activity only on proliferative MDA-MB-231 cell line whereas only a slight toxicity was detected on non-proliferative SUM1315 cell line model. These results suggest that proliferative-3D-cell-culture models represent a great tool for antimitotic drugs evaluation.

Epirubicin is known to be more effective on triple-negative breast tumors [28]. This drug is a very powerful non cell-cycle dependent DNA intercalating agent from the anthracycline family. With epirubicin, cytotoxicity tests profiles with both cell lines were different. Indeed, this drug showed a dose-dependent cytotoxic activity, whatever cell culture conditioning regimens. Nevertheless, 3D cell culture remained less sensitive to epirubicin than 2D cell culture. These data suggest that the non-proliferative and proliferative-3D-cell-culture models may be adapted to the evaluation of conventional cytotoxic drugs.

All these results demonstrated firstly a difference in response between 2D and 3D cell culture models for the two TNBL cell lines. Indeed, a systematic lowered sensitivity was detected for 3D cell culture compared to monolayer cell culture probably due to 3D cellular heterogeneity and/or resistance phenotypes. This reproduces partly tumor response *in vivo* [27]. More, according to drug's mechanism of action, *i.e.* cytotoxic or cytostatic, the choice of a proliferative or non-proliferative 3D model seem to be decisive for drug efficacy's prediction. These observations underline the added value of using 3D cell culture as a tool to explore sensitivity and resistance to chemotherapeutics, among the arsenal of models currently used in preclinical studies.

## MATERIALS AND METHODS

### 2D cell culture

BLTN cell lines SUM1315 and MDA-MB-231 were obtained from Asterand. SUM1315 cells were seeded at 25 000 cells per mL in 75 cm$^2$ culture dish (Falcon®) at 37°C under 5% $CO_2$, in 15 mL Ham's F12 medium (Gibco®) supplemented with 5% decomplemented fetal calf serum, 10 mM HEPES* buffer, 20 mg/mL gentamycin, 10 ng/mL EGF and 4 μg/mL of insulin, according to the supplier's instructions [29, 30]. The MDA-MB-231 cell line was seeded at 25 000 cells per mL in 75 cm$^2$ culture dish (Falcon ®) at 37°C under 5% $CO_2$, in 15 mL RPMI 1640 medium (Gibco®) supplemented with 10% decomplemented fetal calf serum and 20 mg/mL gentamycin, according to the supplier's instructions [29, 30].

Cells having reached confluence were washed with Phosphate Buffer Saline (PBS, 1X, Sigma®) and trypsinized (Trypsin 1X, Sigma®) for 5 min at 37°C. Cells were taken up in 12 mL of culture medium and centrifuged for 10 minutes at 250 G. The pellet was taken up in 5 mL of culture medium and the number of cells per mL was determined by a cell count using a vital stain Trypan Blue. This allowed for counting the cell dilutions necessary for seeding cells in varying concentrations and defined for each experiment.

### Liquid overlay 3D cell culture

Cells were seeded in "Ultra-Low-Attachment" round bottom microplates (Corning®) in order to prevent cellular adhesion to the support. After 24h of incubation, Geltrex® was deposited on aggregated cells, and microplates were agitated on a microplate shaker at 185 rpm for 20 min. Spheroid formation and diameter measurements were assessed with a light microscope (CFM-BDS-200 Realux Cofemo) associated with a camera (Industrial Digital Camera, UCMOS05100KPA) and ToupView® software.

### Metabolism activity assessment using the resazurin test

The resazurin test is a colorimetric assay for cell viability that involves the reduction of resazurin to resorufin by the metabolically active mitochondria of cells. Resorufin is a fluorescent compound that exhibits a pink colouring. After a time of incubation, the amount of resorufin formed is directly proportional to the number of metabolically active cells. Cells cultured in 2D or 3D treated with the drugs for 5 days at 37°C were washed with PBS* and incubated with 60 μM of resazurin solubilized in PBS. Optical densities were measured at 570 and 620 nm after 6 to 15 hours of incubation. The amount of reduced compound (resorufin) corresponding to the number of active cells per well was calculated for each condition. The survival rate was determined considering the ratio ($OD_{570-620nm}$ treated well/$OD_{570-620nm}$ untreated control) x 100. The half maximal inhibitory concentration ($I_{C50}$) was calculated with the following formula: $I_{C50}$ = EXP (LN (concentration > 50% inhibition) - ((signal > 50% inhibition - 50) / (signal > 50% inhibition - signal < 50% inhibition) * LN (concentration >50% inhibition / concentration <50% inhibition))).

### Imaging of 3D cell culture with the Live/Dead® kit

The Live/Dead® kit (Viability/Cytotoxicity kit, Molecular Probes™) is composed of two fluorochromes, the calcein-AM (green fluorescence) which is retained in viable cells and the ethidium homodimer-1 (ethD1, red fluorescence) which penetrates and binds to DNA of cells in phase of apoptosis or necrosis. The two probes were diluted in phosphate saline buffer to final concentrations of 4 mM for Calcein-AM and 2 mM for ethD-1. After several washings in Phosphate Buffer Saline (PBS), spheroids were incubated with 100 μL of the Calcein-AM/EthD-1 solution for at least 45 min., protected from light. Observations were made with an epifluorescence microscope (Leica DMI 3000B, FITC canal) associated with CCD camera and Leica® Software.

### Scanning electron microscopy

Scanning Electron Microscopy (SEM) assessed organization of cell components of the spheroids. Spheroids were washed with 0.2 M sodium cacodylate buffer pH 7.4 and fixed in 1.6% glutaraldehyde in cacodylate buffer overnight at 4°C. They were then rinsed in the same buffer and post-fixed for 1h with osmic acid in cacodylate buffer at ambient temperature. For spheroid visualization, a solution of 0.15% ruthenium red in water was added to the washes and fixative solutions. Fixed spheroids were then progressively dehydrated using a graded ethanol series (10 min. each in 25%, 50%, 70%, 95%, 100% ethanol) and hexamethyldisilazane (3x10 min.). After drying, the samples were mounted on stubs using adhesive carbon tabs and sputter-coated with gold-palladium (JFC-1300, Jeol). Spheroids were observed with a scanning electron microscope Jeol 6060-LV at 5 kV in high-vacuum mode.

## Transmission electron microscopy

Transmission Electron Microscopy (TEM) achieved ultrastructural observations of spheroids. Spheroids were washed in 0.2M sodium cacodylate buffer (pH7.4), fixed in 1.6% glutaraldehyde for 24 hours at 4°C. Specimens were then washed three times (10 min.) in Na cacodylate buffer, post fixed 1h with 1% osmium tetroxide in Na cacodylate buffer and washed three times in Na cacodylate buffer. In order to visualize the spheroids, a solution of 0.15% ruthenium red in water was added to the buffer and fixative solutions. Fixed spheroids were then progressively dehydrated in ethanol (70%, 95%, 100%) and acetone. Subsequently, they were infiltrated with acetone and EPON resin mixture (2:1, 1:1, 1:2 for 1h). Specimens were embedded in resin EPON overnight at room temperature, and cured 2 days at 60°C. Thin sections (70 nm) were cut using a UC6 ultramicrotome (Leica) and stained with uranyl acetate and Pb citrate. Carbon was evaporated using CE6500 unit. Observation of the ultrastructural organization of cells within the spheroids was then conducted with a transmission electron microscope (Hitachi H-7650) at 80 kV acceleration voltage. Micrographs were made using a Hamamatsu AMT camera placed in a side position.

## Drug solubilization and cell exposure

Docetaxel, cisplatin and epirubicin were prepared at a stock solution of 10 mM. Cisplatin and docetaxel were not soluble in water. These were solubilized in dimethylsulfoxide (DMSO) and the final concentration in each cell culture well was always of 0.1%. Epirubicin was solubilized in distilled water. For all drug treatment experiments, cells cultured in 2D and 3D condition were seeded at a concentration of 1000 cells per well for both cell lines. Cells were then exposed to the drugs 24h after seeding for 2D cell culture (phase of exponential growth) and 24h after Geltrex® adding for 3D cell culture. After five days of incubation, cell viability was then assessed for each condition with the resazurin test. For 3D cell culture, spheroid diameter measurements and Live/Dead® viability/mortality tests were also performed at the end of each treatment.

## Statistical analysis

Results were expressed as means ± standard deviation of $n$ independent experiments. All experiments were performed at least in triplicate and then statistically compared using a Student's t-test. Tests were two-sided and the nominal level of significance was $p<0.05$ (*), $p<0.01$ (**), $p<0.001$ (***), $p<0.0001$ (****) and $p<0,00001$ (*****).

## Abbreviations

HEPES: *4*-(2-hydroxyethyl)-1-piperazineethanesulfonic acid

PBS: phosphate buffered salin

## Author contributions

C.D., R.D., M.B., P.D., E.M., C.A., F.P.L. conceived and designed the study; C.D., R.D., C.B., C.S. performed the experimental work; C.D., R.D., P.D., C.B. E.M. and M.B. analyzed the data; C.D. performed statistical analysis; C.D. R.D., C.B. and M.B. wrote the manuscript. All authors reviewed the manuscript.

## ACKNOWLEDGMENTS

Our thanks to Lisa Prophète for the technical work, Fabrice Kwiatkowski for his advice with the statistical analysis and Lorraine Novais Gameiro for her assistance with TEM/SEM technologies at the "Centre Imagerie Cellulaire Santé" (Clermont Auvergne University).

## CONFLICTS OF INTEREST

The authors declare no competing financial interests.

## FUNDING

These works were supported by the Regional Anticancer League (*La Ligue Régionale Contre le Cancer*) and solidarity sport events organized by "La Chamaliéroise" and "La Clermontoise", two associations involved in the fight against cancer.